\documentclass[a4paper,11pt,fleqn]{article}
\usepackage{amsfonts}

\def\nn{\nonumber}
\def\non{\nonumber\\}
\def\be{\begin{equation}}
\def\ee{\end{equation}}
\def\ben{\begin{displaymath}}
\def\een{\end{displaymath}}
\def\ba{\begin{eqnarray}}
\def\ea{\end{eqnarray}}

\def\d{\delta}
\def\e{\varepsilon}
\def\f{\varphi}

\def\g{\gamma}
\def\G{\Gamma}

\def\l{\lambda}

\def\r{\rho}

\def\t{\tau}

\def\e{\epsilon}

\def\H{{\cal H}}

\def\cP{{\cal P}}

\def\R{\mathbb{R}}

\def\la{\label}
\def\ci{\cite}

\def\Ref#1{(\ref{#1})}

\def\f{\frac}
\def\ft#1#2{{\textstyle {\frac{#1}{#2}} }}
\def\i{\infty}
\def\p{\partial}

\def\tr{{\rm tr}}

\begin{document}
\renewcommand{\thefootnote}{\fnsymbol{footnote}}
\begin{flushright}
{\small AEI-036 \\
gr-qc/9705013 \\
May 1997}
\end{flushright}
\begin{center}
{\bf\LARGE  Canonical Quantization of}\medskip\\
{\bf\LARGE Cylindrical Gravitational Waves}\medskip\\ 
{\bf\LARGE with Two Polarizations}\bigskip\\ 
{\large D. Korotkin$^{1,}$\footnote{On leave of absence from
    Steklov Mathematical Institute, Fontanka, 27, St.Petersburg 191011
    Russia} and~ H. Samtleben$^2$\bigskip\medskip\\ }
{ $^1$ II. Institut f\"ur Theoretische
Physik, Universit\"at Hamburg,\\  Luruper Chaussee 149, D-22761
Hamburg, Germany}\\ \small E-mail: korotkin@x4u2.desy.de\medskip\\
{ $^2$ Max-Planck-Institut f\"ur Gravitationsphysik,
Albert-Einstein-Institut,\\
Schlaatzweg 1, D-14473 Potsdam, Germany}\\ \small E-mail:
henning@aei-potsdam.mpg.de 
\end{center}

\renewcommand{\thefootnote}{\arabic{footnote}}
\setcounter{footnote}{0}
%\vspace*{1.3cm}
\vspace*{0.6cm}
\hrule
\begin{abstract}
\noindent
The canonical quantization of the essentially nonlinear
midisuperspace mo\-del describing cylindrically symmetric
gravitational waves with two polarizations is presented. A Fock space
type representation is constructed. It is based on a complete set of
quantum observables. Physical expectation values may be calculated in
arbitrary excitations of the vacuum. Our approach provides a
non-linear generalization of the quantization of the collinearly
polarized Einstein-Rosen gravitational waves.
\end{abstract}
\medskip
\hrule
\vspace{1cm}
%\newpage

The quantization of dimensionally reduced models of 4d Einstein
gravity serves as interesting testing ground for many issues of
quantum gravity. The physical output of this approach to an
understanding of characteristic features of the full theory however
strongly depends on the complexity of the model under consideration.

The probably simplest and best understood examples are the
mini-super\-space models \cite{mini} which contain only a finite number
of physical degrees of freedom and thus hide the field effects of
quantum gravity.  A more complicated example of steady interest is
given by the midi-superspace model of cylindrically symmetric
gravitational waves with one polarization \ci{Kuch71,AshPie96a}. This
model already involves an infinite number of degrees of freedom. It
becomes treatable with the methods of flat space quantum field theory,
because the Einstein field equations essentially reduce to the
axisymmetric 3d Laplace equation.  This underlying linearity on the
other hand may conceal typical nonlinear features of quantum gravity.

It is the purpose of this letter to generalize the results of
\ci{Kuch71,AshPie96a} to cylindrical gravitational waves with two
polarizations, where the Einstein equations become truly
nonlinear. We achieve a consistent canonical quantization in terms of
a complete set of quantum observables. Creation and annihilation
operators are identified in a kind of Fock space representation of
these observables. The full Hilbert space of physical quantum states
is build from excitations of the vacuum. The presented techniques
allow to calculate all physical expectation values in arbitrary
quantum states.

We start from a general space-time with cylindrical symmetry,
i.e.~assume existence of two commuting Killing vector fields, one of 
which has closed orbits. Choose coordinates such that the Killing
vector fields are given by $\p_z$ and $\p_\varphi$ associated to the
axis of symmetry $z$ and the azimuthal angle $\varphi$
respectively. Further gauge-fixing brings the metric into the standard
form \ci{Komp58}   
\be
ds^2 ~=~   e^{\G(\r,\t)} (-d \t^2 + d\r^2) +\r\,g_{ab} (\r,\t) dx^a dx^b
\qquad a,b=2,3 ~, \la{m}
\ee
with $x^2\!\equiv\!z$, $x^3\!\equiv\!\varphi$, radial coordinate $\r$
and time $\t$. The symmetric $2\!\times\!2$ matrix $g$ is restricted
by  the condition ${\rm det}\;g \!=\!1$.

The Einstein field equations consist of two parts: the Ernst equation
for the matrix $g$:
\be
\p_\r(\r g^{-1} \p_\r g ) +\p_\t(\r g^{-1}\p_\t g ) ~=~0 ~,
\la{ee}\ee
and the equation for the conformal factor $\G$:
\be
\la{cf}
\G(\r,\t) ~=~ \ft12\int_0^\r \r'\,d\r'\; 
\tr \Big((g^{-1} \p_{\r'} g )^2 + (g^{-1}\p_\t g )^2\Big) ~.
\ee
The conformal factor at spatial infinity $\G_\i\equiv\G(\r\!=\!\i)$
generates evolution with respect to the time coordinate $\t$. Its
exponential measures the total energy per unit length in $z$-direction
and the deficit angle in the asymptotic region 
\be\la{dah}
H^t ~=~ \frac1{\pi G} \,\varphi_0 ~=~ 
\frac2{G}\,(1 - e^{-\G_\i/2})~.
\ee 

The reduction of the metric to the form \Ref{m} can be performed
within the canonical formalism, such that the Poisson bracket of
the reduced model is the Dirac bracket of the original structure after
appropriate gauge fixing \ci{AshPie96a}. 
The resulting canonical Poisson structure is easily extracted from
the effective Lagrangian density ${\cal L}^{(2)}$ that comes from
reduction via Killing symmetries and gauge fixing of the original
Lagrangian ${\cal L}^{(4)}_{EH}\!=(1/G)\!\sqrt{|g_{\mu\nu}|}R^{(4)}$: 
\ben
{\cal L}^{(2)}(\r,\t) ~=~  \f1{2G}\,\r\,
\tr \Big((g^{-1}\p_\r g )^2 - (g^{-1}\p_\t g )^2\Big)~.
\een
In matrix components $g_{ab}$, the Poisson brackets read
\ben
\Big\{g_{ab}(\r), (g^{-1} \p_\t g g^{-1})_{cd}(\r') \Big\} 
~=~\frac{G}{\r}\,\d_{ad}\d_{bc}\d(\r-\r')~.
\een
The restrictions of symmetry and unit determinant of $g$ require some
additional technical effort and have been taken into account
in the derivation of the following results.

{\it Collinear polarizations.} Among the simplest nontrivial metrics
of this model are the collinearly polarized gravitational waves
discovered by Einstein and Rosen \ci{EinRos37}. They correspond to a
diagonal form of the matrix $g\equiv{\rm diag}(e^\phi,\,e^{-\phi})$,
i.e.~the number of degrees of freedom reduces to one. The Ernst
equation \Ref{ee} in this case reduces to the cylindrical Laplace
equation   
\ben
-\p^2_\t\phi+\r^{-1}\p_\r\phi + \p^2_\r\phi=0 ~,
\een
with general solution
\ben
\phi(\r,\t)=\int_0^\i \left[A_+(\l)J_0 (\l\r) e^{i \l \t} + 
A_-(\l)J_0 (\l\r) e^{-i \l \t}\right] d\l ~,
\een
where $J_0$ denote Bessel functions of the first kind. The coefficients
$A_+\!=\!\overline{A_-}$ build a complete set of observables with
canonical Poisson brackets 
\be
\Big\{A_+(\l),\; A_- (\l')\Big\} = G\,\delta (\l-\l')~.
\la{psA}\ee
Thus, quantization of this structure is straightforward \ci{AshPie96a}
and gives rise to a representation in terms of creation and
annihilation operators 
\be\la{fock}
A_- |0\rangle = 0 \qquad {\rm with} \qquad A_+=A_-^\dagger ~. 
\ee
In particular, coherent quantum states may be constructed in the same
way as in flat space quantum field theory. Recent discussion however
has shown, that these states do not provide coherence of all essential
physical quantities \ci{Asht97}. 

As the first step towards the general case, we cast the truncated
model of collinear polarization into a form that will allow proper 
generalization. Introduce new variables 
\be
T_\pm (w) ~\equiv~ \exp\int_0^\i A_\pm (\l) e^{\pm iw \l} d\l~,
\la{Tpma}\ee
which build an equivalent complete set of observables. In the Fock
space representation \Ref{fock} $T_-(w)$ is represented as identity,
whereas $T_+(w)$ generates the coherent state associated to a
classical field that on the symmetry axis $\r\!=\!0$ is peaked as a
$\d$-function at $\t_0\!=\!w$. In terms of these new variables, the
Poisson structure \Ref{psA} becomes
\be
\Big\{T_- (v), T_+ (w)\Big\} ~=~ -\f{G}{v - w}\,T_-(v) T_+(w) ~.
\la{TTa}\ee
We shall see in the sequel, that it is this quadratic form of Poisson
brackets which generically appears in the case of two
polarizations. Linearization to \Ref{psA} is a special feature of the
truncated model but not possible in the general case.

{\it Two polarizations.} 
In general, the Ernst equation \Ref{ee} does not admit explicit
solution. However, it is possible to construct the analogue of the
quantities $T_\pm$ defined above. Inspired by the auxiliary linear
system associated to the Ernst equation \ci{BelZak78} we define for
real $w$  
\be\label{T}
T_\pm(\t, w) ~\equiv~ \lim_{\e\rightarrow0} \,\left\{ 
\cP \exp \int_0^\infty \!\!  d\rho\;2\left(
\frac{\g_\pm^2\;g^{-1}\p_\r g}{1-\g_\pm^2}-
\frac{\g_\pm\;g^{-1}\p_\t g}{1-\g_\pm^2}\right) \!\right\} ,
\ee
with 
\ben
\g_\pm ~\equiv~ 
-\frac1{\rho}\Big(
w\!\pm\!i\e\!-\!\t+\sqrt{(w\!\pm\!i\e\!-\!\t)^2-\rho^2}\Big) ~.
\een
For diagonal $g$, this definition indeed reduces to \Ref{Tpma} above.
The variables $T_\pm$ are still constants of motion, 
i.e.~$\p_\t T_\pm(\t,w) = 0$. They turn out to admit holomorphic
expansion into the upper and lower half of the complex plane
respectively. Definition \Ref{T} further implies $\det T_\pm\!=\!1$
and $T_+=\overline{T_-}$. 

As another important result, the matrix product $M=T_+T_-^t$ coincides
with the values of the metric $g$ on the symmetry axis:
\begin{equation}
  \label{M}
M(w)~\equiv~ T_+(w) T_-^t(w) ~=~ g(\rho\!=\!0, \t\!=\!w) ~.
\end{equation}
In particular, it is symmetric and real:
\be\la{con}
M(w) = M^t(w) \qquad {\rm and } \qquad  
M(w) = \overline{M(w)}  ~.
\ee
Since the $T_\pm$ contain the initial values of the metric on the
symmetry axis, they contain sufficient information to
recover $g$ everywhere by means of \Ref{ee} (note that $\p_\rho
g(\r\!=\!0)=0$). Thus, the set of $T_\pm(w)$ builds a {\it
complete} set of observables for the Ernst equation.

Continuing the program of canonical quantization we next calculate
their Poisson algebra to subsequently quantize it. A direct but
lengthy calculation reveals a {\it quadratic} Poisson algebra for the
matrix entries $T^{ab}_\pm(w)$: 
\ba 
\Big\{T^{ab}_\pm(v),T^{cd}_\pm(w)\Big\} &=& 
\frac{G}{v-w}\:
\Big(T^{ad}_\pm(v)T^{cb}_\pm(w)-T^{cb}_\pm(v)T^{ad}_\pm(w)\Big)
~,\la{pa1}\\ 
\Big\{T^{ab}_-(v),T^{cd}_+(w)\Big\} &=& 
\frac{G}{v-w}\:
\Big(T^{ab}_-(v)T^{cd}_+(w)-T^{cb}_-(v)T^{ad}_+(w) \la{pa2}\\
&&\hspace{3.7em}{}-\d^{bd}\,T^{am}_-(v)T^{cm}_+(w)
\Big)~.\nn
\ea
which consistently encloses the scalar algebra \Ref{TTa} in the
components $T_\pm^{11}(w)$.  
Quantization of this quadratic structure is rather more subtle than
that of a linear algebra, since there appear obvious ambiguities on
the r.h.s.~due to different orderings of the quadratic
expressions. 
Fortunately, the proper quantum analogue of the Poisson brackets
\Ref{pa1} is known in the theory of integrable systems
\ci{Skly82} as the so-called $\mathfrak{sl}(2)$-Yangian 
algebra   
\be \la{y1}
\Big[T^{ab}_\pm(v),T^{cd}_\pm(w)\Big] ~=~ 
\frac{i\hbar G}{v-w}\:
\Big(T^{cb}_\pm(w)T^{ad}_\pm(v)-T^{cb}_\pm(v)T^{ad}_\pm(w)\Big)~.
\ee
The proper quantization of \Ref{pa2} leads to a set of mixed relations
\ba 
\Big[T^{ab}_-(v),T^{cd}_+(w)\Big] &=& 
\frac{i\hbar G}{v-w+2i\hbar G}\:
\Big(T^{ab}_-(v)T^{cd}_+(w)
-T^{cb}_-(v)T^{ad}_+(w)\Big) \non
&&{}-\frac{i\hbar G}{v-w}\:\d^{bd}\,T^{cm}_+(w)T^{am}_-(v)~.\la{y2}
\ea
The shift of the denominator on the r.h.s.~provides quantum
corrections of \Ref{pa2} of higher order in $\hbar$ which are
necessary  for compatibility of these commutation relations with the
quantum analogue of the symmetry \Ref{con}: 
\be\la{qcon}
T_+(w)T_-^t(w) = \Big(T_+(w)T_-^t(w)\Big)^t ~.
\ee
Again, the ordering of these quadratic expressions is now
essential. Classically, $M(w)$ contains the essential physical
objects according to \Ref{M}. In the quantum model, the definition
$M(w)=T_+(w)T_-^t(w)$ ensures, that the commutation relations
\Ref{y1}, \Ref{y2} actually yield a {\it closed} commutator algebra of
the matrix entries of $M(w)$. Moreover, these are hermitean operators,
provided that 
\be\la{herm}
T_+^{ab}(w) = \Big(T_-^{ab}(w)\Big)^\dagger  \qquad w\in \R~,
\ee
in accordance with the classical relations.  Finally, the classical
condition of unit determinant $\det T_\pm(w)=1$ requires quantum
corrections because of the nonlinear terms and is substituted by the
``quantum determinant'' \ci{IzeKor81}
\be\la{qdet}
T_\pm^{11}(w\!+\!i\hbar G)T_\pm^{22}(w)-
T_\pm^{12}(w\!+\!i\hbar G)T_\pm^{21}(w)~=~1~,
\ee
which is indeed compatible with the relations \Ref{y1}, \Ref{y2} and
may as such be imposed as an operator identity. 

Summarizing, we have formulated the consistent quantum model in terms
of the operators $T_\pm^{ab}(w)$, subject to the commutation relations
\Ref{y1}, \Ref{y2}, as well as to unit quantum determinant \Ref{qdet},
hermiticity \Ref{herm} and symmetry \Ref{qcon}. We are now in position
to introduce a Fock space type representation of this algebra,
inspired by the scalar case \Ref{fock}. Let therefore $T_-(w)$ act
trivially on the vacuum
\be\la{gfock}
T_-^{ab}(w) |0\rangle ~=~ \d^{ab} |0\rangle~, 
\ee
and $T_+(w)$ generate the Hilbert space of physical states
\ben
\H = \left\{ \left.\left(\prod_{i=1}^{m} 
T_+^{a_ib_i}(w_i)\right)|0\rangle \quad\right|
\,m,a_i,b_i,w_i \;\right\}, 
\een
where all the excitations are not independent but obey the relations
\Ref{y1}, \Ref{qcon} and \Ref{qdet} for $T_+$.  The intuitive idea
that the $T_+^{ab}(w)$ generate the complete spectrum of states is not
only supported by the exactly solved scalar case from above, but even
stronger by the fact, that the conserved charges $T_+(w)$ canonically
Poisson-generate the Geroch group \ci{KorSam96a} which as a symmetry
group acts transitively among the classical solutions of the field
equations \ci{Gero72}.

It is straightforward to further extract all relevant physical
information from the quantum model. The hermiticity relations
\Ref{herm} together with the commutation relations \Ref{y2} allow to
calculate the expectation values of arbitrary polynomials in the
$T_\pm^{ab}(w)$ in arbitrary excitations of the vacuum. Indeed, a
closer look at \Ref{y2} shows, that by means of these relations, the
$T_-^{ab}(w)$ may be shuffled through to the right in any sequence of
operators, where they finally ``annihilate'' the vacuum according to
\Ref{gfock}. As an illustration we state the scalar product between
excitations of the first level $|T^{ab}_+(w)\rangle\equiv
T^{ab}_+(w)|0\rangle$:
\ba
\langle T^{ab}_+(v)|T^{cd}_+(w) \rangle &=& 
\langle 0|T^{ab}_-(v)T^{cd}_+(w) |0\rangle\\
&=& \d^{ab}\d^{cd} + \frac{i\hbar G}{v-w}\,
\Big(\d^{ab}\d^{cd}-\d^{ad}\d^{bc} -\d^{ac}\d^{bd} \Big)~.\nn
\ea

We can also derive expectation values of the conformal factor $\G_\i$
and its exponential $e^{\G_\i}$, related to energy, deficit angle
and metric components at infinity \Ref{dah}. Namely, it may be shown
that classically    
\be\la{Hcom}
\{\G_\i, T_\pm(w)\} ~=~ G\,\p_wT_\pm(w)~.
\ee
In the quantum theory, the conformal factor can thus be represented as
derivation operator $i\hbar G\,\p/\p w$, such that its exponential
$e^{\G_\i}$ becomes the shift operator $w\mapsto w+i\hbar G$. It is
then an elementary exercise to calculate its matrix elements between
arbitrary states of $\H$.

The presented quantum model provides the exact quantization of a
midi-super\-space model of quantum gravity with essential nonlinear
characteristics. The complete set of quantum observables and
the complete spectrum of physical quantum states are at hand. The
techniques are sufficiently developed to start exploring the
properties of the spectrum and relevant observables.  

It would be of high interest to identify some kind of coherent states
in this model, i.e.~quantum states with certain semi-classical
properties. Due to the nonlinear setting it is reasonable to
suspect, that not all the standard properties of usual coherent states
can be satisfied. The fact, that the traditional framework of coherent
states may be too restrictive for the description of quantized
gravitational waves is actually supported by recent observations in
the linear model \ci{Asht97}.

Another exciting feature of this quantum model emerges from the quantum
analogue of the determinant \Ref{qdet}: In view of the physical
interpretation of $M(w)$ \Ref{M}, which supplies the spectral parameter
$w$ with a space-time meaning, it is tempting to consider \Ref{qdet}
as a sign of arising nonlocality of the quantum operators on the
Planck scale.

Since the presented quantization mainly employs the group-theoretical
properties of the model, it will allow natural generalization to other
and more complicated models of dimensionally reduced gravity,
including higher dimensional supergravity as well as Einstein-Maxwell
systems. Similarly it should find application to the Gowdy
model, where $\rho$ becomes a time-like variable \ci{Mena97}.  The
weak field limit of the nonlinear Poisson structure in this case is
isomorphic to the isomonodromic Poisson structure quantized in
\ci{NiKoSa96}. With different norm of the reducing Killing vector
fields, the whole scheme may furthermore be applied to stationary
axisymmetric spacetimes, providing an exact quantization of the black
hole solutions in a vast class of models.

We would like to thank H. Nicolai and V. Schomerus for valuable
discussions. The work of D.K. was supported by DFG Contract Ni
290/5-1. H.S. thanks Studienstiftung des deutschen Volkes for financial
support.


\begin{thebibliography}{10}

\bibitem{mini} 
See e.g.
A. Ashtekar, R. Tate, and C. Uggla, {\em Int. J. Mod. Phys.} {\bf D2},
15 (1993);
H. Kastrup and T. Thiemann, {\em Nucl. Phys.} {\bf B425}, 665 (1994);
D. Marolf, {\em Phys. Rev.} {\bf D53}, 6979 (1996);
G. A. Mena Marug{\'a}n, {\em Phys. Rev.} {\bf D53}, 3156 (1996).

\bibitem{Kuch71}
K. Kucha\v{r}, {\em Phys. Rev.} {\bf D4}, 955  (1971).

\bibitem{AshPie96a}
A. Ashtekar and M. Pierri, {\em J. Math. Phys.} {\bf 37}, 6250 (1996).

\bibitem{Komp58}
A. Kompaneets, {\em Sov. Phys. JETP} {\bf 7}, 659 (1958).

\bibitem{EinRos37}
A. Einstein and N. Rosen, {\em J. Franklin Inst.} {\bf 223}, 43 (1937).

\bibitem{Asht97}
A. Ashtekar, {\em Phys. Rev. Lett.} {\bf 77}, 4864 (1996);
R. Gambini and J. Pullin, Preprint CGPG-97/4-1, gr-qc/9703088 (1997).

\bibitem{BelZak78}
V. Belinskii and V. Zakharov, {\em Sov. Phys. JETP} {\bf 48}, 985
(1978); 
D. Maison, {\em Phys. Rev. Lett.} {\bf 41}, 521 (1978).

\bibitem{Skly82}
L. Faddeev, E. Sklyanin, and L. Takhtajan, {\em Theoret. Math. Phys.}
{\bf 40}, 194 (1979); 
L. Faddeev, N. Reshetikhin, and L. Takhtajan, {\em Leningrad Math. J.}
{\bf 1}, 193 (1990). 

\bibitem{IzeKor81}
A. Izergin and V. Korepin, {\em Sov. Math. Doc.} {\bf 26}, 653 (1981).

\bibitem{KorSam96a}
D. Korotkin and H. Samtleben, Preprint DESY-96-245, gr-qc/9611061, 
{\em Class. Quantum Grav.} accepted for publication.
To be precise, the classical Geroch group is generated  by the
operators $T_+^{-1}(w)\,{\rm ad}_{T_+(w)}\,$, which as a corollary of
\Ref{pa1} build half of the affine algebra
$\widehat{\mathfrak{sl}_2}$. 


\bibitem{Gero72}
R. Geroch, {\em J. Math. Phys.} {\bf 13}, 394 (1972).


\bibitem{Mena97}
A recent treatment of the Gowdy model in terms of Ashtekar
variables has been given in
V. Husain, {\em Phys. Rev.} {\bf D53}, 4327 (1996);
G. A. Mena Marug{\'a}n, Report No. gr-qc/9704041 (1997).

\bibitem{NiKoSa96}
D. Korotkin and H. Nicolai, {\em Nucl. Phys.} {\bf B475}, 397 (1996);
D. Korotkin and H. Samtleben, Preprint DESY-96-130, hep-th/9607095,
{\em Commun. Math. Phys.} accepted for publication.  
\end{thebibliography}
\end{document}